\documentclass[pdflatex,aps,twocolumn,superscriptaddress,prl,fleqn,showpacs,nofootinbib,preprintnumbers]{revtex4}
\usepackage{amssymb,amsmath,epsfig}

\renewcommand{\d}{\mathrm{d}}

\renewcommand{\d}{\mathrm{d}}

\newcommand{\bea}{\begin{eqnarray}}
\newcommand{\eea}{\end{eqnarray}}

\newcommand{\nn}{\nonumber \\}
\newcommand{\nnn}{\nonumber }

\def\slash#1{\setbox0=\hbox{$#1$}  
   \dimen0=\wd0     
   \setbox1=\hbox{/} \dimen1=\wd1  
   \ifdim\dimen0>\dimen1   
      \rlap{\hbox to \dimen0{\hfil/\hfil}} 
      #1     
   \else     
      \rlap{\hbox to \dimen1{\hfil$#1$\hfil}} 
      /      
   \fi}      %

\begin{document}

\title{A partonic description of the transverse target single-spin asymmetry in inclusive DIS}

\author{Marc Schlegel}
\email{marc.schlegel@uni-tuebingen.de}
\affiliation{Institute for Theoretical Physics,
                Universit\"{a}t T\"{u}bingen,
                Auf der Morgenstelle 14,
                D-72076 T\"{u}bingen, Germany}

\begin{abstract}
The single-spin asymmetry of unpolarized leptons scattering deep-inelastically off transversely polarized nucleons is studied in a partonic picture within a collinear twist-3 framework. Since this observable is generated by multi-photon exchanges between lepton and nucleon a partonic description of the asymmetry contains essential elements of a full next-to-leading order calculation for single-spin asymmetries in perturbative Quantum Chromodynamics. In particular it is shown how nontrivial cancellations between kinematical and dynamical twist-3 contributions lead to a well-behaved and finite formula. This final result can be expressed in terms of multipartonic quark-gluon and quark-photon correlation functions. Hence, a measurement of the transverse target single-spin asymmetry may provide new constraints on these multipartonic correlations.
\end{abstract}

\date{\today}

\maketitle

{\it I. Introduction \---}
One of the most fundamental and basic processes in hadronic physics is the deep-inelastic scattering (DIS) of leptons off nucleons, $\mathrm{l}(l)+\mathrm{N}(P)\to \mathrm{l}(l^\prime)+\mathrm{X}$. Historically, the experimental and theoretical analyses of this simple reaction has provided valuable information on the partonic structure of the nucleon and has led to important progress in the understanding of the strong interactions \--- and ultimately to the developement of Quantum Chromodynamics (QCD). 
If an exchange of a single virtual photon between lepton and nucleon is assumed one can decompose the differential DIS cross section into four structure functions, commonly denoted as $F_1$, $F_2$, $g_1$, $g_2$. Two of them ($F_1$, $F_2$) describe the scattering of unpolarized leptons and nucleons while the others ($g_1$, $g_2$) describe processes with both particles being polarized. All of these structure functions have been studied intensely over the last decades. In particular the observation of Bjorken scaling within the parton model and the explanation of its logarithmic violation has been considered as a big success of perturbative QCD. Measurements of the polarized $g_1$ structure function have led to the surprising insight that only a small fraction of the nucleon spin is carried by quark spins \--- an observation sometimes referred to as the 'spin crisis'. On the other hand the structure function $g_2$ has played a central role in transverse spin physics and in the understanding of 'higher-twist' observables.

In principle, single-spin observables in inclusive DIS with either the lepton or nucleon being transversely polarized are equally fundamental. They have received less attention because single-spin observables strictly vanish due to time-reversal invariance for a single photon exchange~\cite{Christ:1966zz}. This argument fails if two (or more) photons are exchanged between lepton and nucleon. Naturally, single-spin effects are expected to be small since exchanges of more than one photon are suppressed by powers of the fine structure constant $\alpha_{\mathrm{em}}=1/137$. 

Experimentally, the single-spin asymmetry (SSA) for a transversely polarized nucleon, denoted by $A_{UT}$, was already measured in 1970 in the resonance region~\cite{EarlyExp}. A result consistent with zero within an error of about $10^{-2}$ was obtained.
A recent measurement of $A_{UT}$ was performed by the HERMES collaboration~\cite{HERMES}, and again a result consistent with zero was found within an error of about $10^{-3}$. Interestingly, preliminary data taken from (ongoing) precision measurements of $A_{UT}$ at Jefferson Lab seem to indicate a non-zero effect~\cite{JLab}.

A theoretical description of the SSA $A_{UT}$ in a partonic picture requires to deal with two distinctive and complementary physical situations: The exchange of two photons between the lepton and either (i) one {\it single} quark or (ii) two {\it different} quarks.
A first attempt to describe $A_{UT}$ for scenario (i) was made fairly recently in Ref.~\cite{OwnWork1}. 
In this work a 'proof of concept' for a non-zero SSA, generated by a two-photon exchange, was presented. In particular it was found that this observable generically behaves like $M/Q$  where $M$ denotes the nucleon mass, $Q^2=-q^2$, and $q=l-l^\prime$ the 4-momentum transfer to the nucleon. Thus the asymmetry is a power suppressed ('twist-3') observable. 
However, the results of Ref.~\cite{OwnWork1} are not conclusive since contributions from multipartonic non-perturbative quark-gluon correlations were not taken into account. For this reason an uncanceled soft singularity remains in the partonic formula presented in Ref.~\cite{OwnWork1}. While it has been conjectured in Ref.~\cite{OwnWork1} that the inclusion of quark-gluon correlations might cancel the soft singularity the implementation of these twist-3 effects within a partonic picture is non-trivial and has been an open issue so far. In the following the conjecture of Ref.~\cite{OwnWork1} is eventually shown to be true. 
In addition quark mass effects proportional to the transversity distribution $h_1^q(x)$ are also relevant for scenario (i) and have been studied in Ref.~\cite{Afanasev}. 

Very recently a partonic formula has been presented in Ref.~\cite{new} for scenario (ii) where the SSA $A_{UT}$ is described in terms of quark-photon correlation functions. Arguments were given in Ref.~\cite{new} that this scenario may dominate at large $x_B$. Whether or not this is true can only be decided once all contributions from both scenarios (i) and (ii) are known and final data on $A_{UT}$ become available.

The purpose of this paper is to complete the previous work of Ref.~\cite{OwnWork1} and to discuss results for the aforementioned multipartonic quark-gluon correlation effects. For the first time this allows to present a complete and comprehensive formula for the transverse target SSA $A_{UT}$ in inclusive DIS in a partonic picture. 

\begin{figure}[t]
\centering
\psfig{file=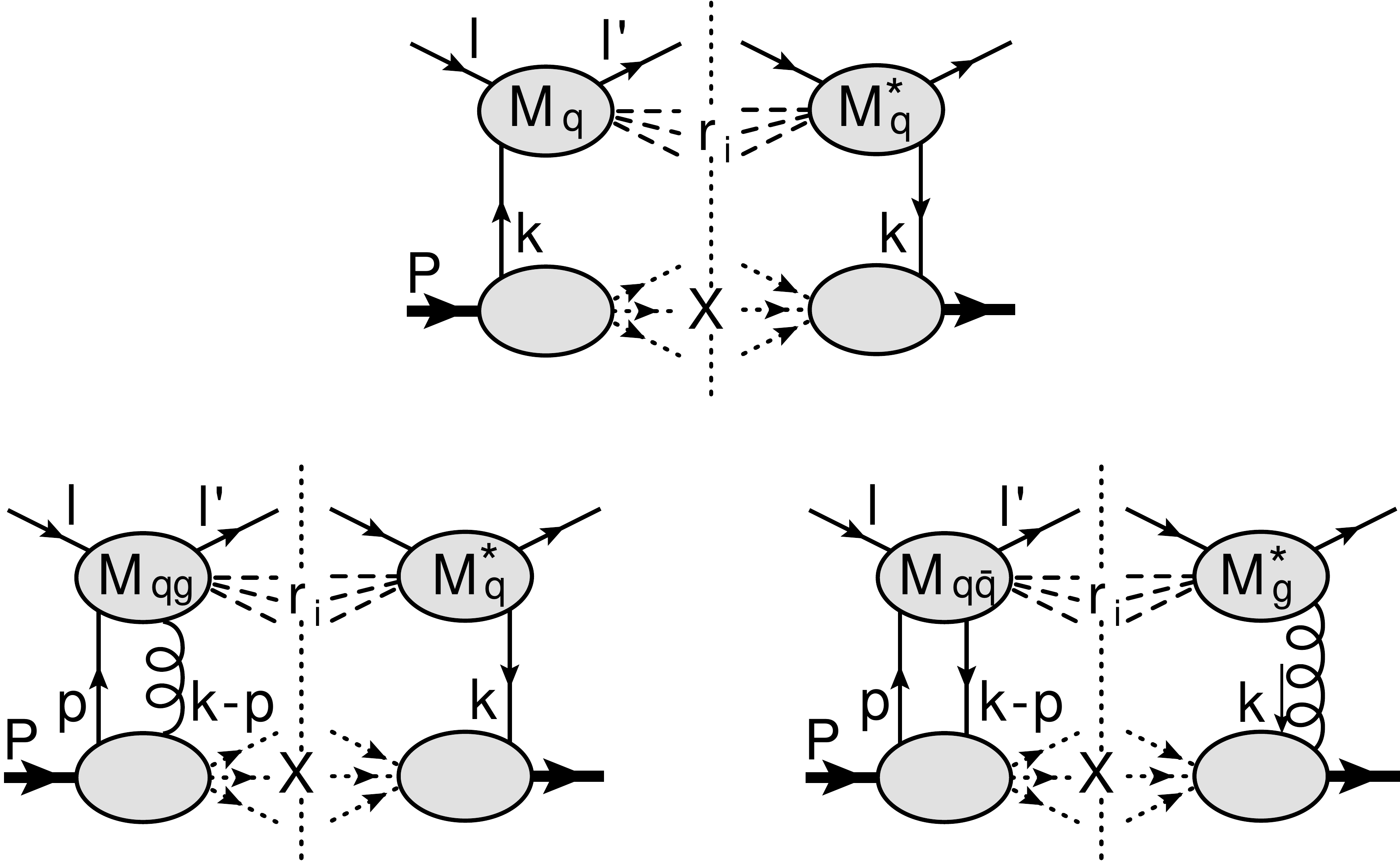, width=0.45\textwidth} 
\caption{ Diagrammatic contributions to the DIS cross section. The upper diagram indicates quark correlations and describes effects of transverse quark motion (first and second line of (\ref{eq:master}), "kinematical twist-3"). The corresonding partonic cross section is labeled by $\d \hat{\sigma}_{q\bar{q}}$. The lower left and right diagrams describe multipartonic quark-gluon correlations ("dynamical twist-3"). The hard parts $\d \hat{\sigma}_{qg\bar{q}}$ (left, third line in (\ref{eq:master})) and $\d \hat{\sigma}_{q\bar{q}g}$ (right, last line in (\ref{eq:master})) are generated from interferences of quark-gluon and quark-antiquark amplitudes with a quark and a gluon amplitude, respectively. The hard factors may contain phase space integrations over unobserved partons carrying momenta $r_i$ in the final state.}
\label{DiagFact}
\end{figure}

{\it II. Twist-3 Formalism \---} The DIS differential cross section can be analyzed in terms of the commonly used DIS variables that are defined as $x_B=Q^2/(2P\cdot q)$ and $y=P\cdot q/P\cdot l$. 
For the description of $A_{UT}$ a transverse (to the lepton plane) spin vector $S_T$ of a polarized nucleon is needed. An azimuthal angle $\phi_s$ between $S_T$ and the lepton plane determines the spatial orientation of $S_T$. 
\\
In order to analyze the transverse target SSA in the parton model it is necessary to apply a formalism that describes not only the leading part of the DIS cross section in an expansion in $M/Q$ but also the subleading part. Diagrammatic factorization as advocated in Ref.~\cite{DiagFact} provides an illustrative way to handle the subleading contributions. The procedure is pictorially sketched in Fig.~\ref{DiagFact} where the diagrammatic separation of the cross section into hard parts and (multipartonic) soft parts is shown. Translating these diagrams in Fig.~\ref{DiagFact} into a formula for the transverse target spin dependent cross section $\d\sigma_{UT}\equiv E^\prime (\d\sigma_{UT})/(\d^{d-1}l^\prime)$ in $d=4+2\varepsilon$ dimensions leads to the following expression (chiral-odd and quark mass effects are neglected),
\begin{eqnarray}
\d\sigma_{UT} & = &- \frac{\pi x_B y M}{4 Q^2} \sum_{q} \int_{x_B}^1 \frac{dx}{x} \Big[
  x g_T^q(x)\ \d \hat{\sigma}_{q\bar{q}}^\perp\label{eq:master}\\
& & - g_{1T}^{(1),q}(x)\ \d \hat{\sigma}_{q\bar{q}}^{\partial,-}
 - f_{1T}^{\perp (1),q}(x)\ \d \hat{\sigma}_{q\bar{q}}^{\partial,+}\nn
& & \hspace{-1cm} -\tfrac{1}{2}\int_0^1\d x^\prime\Big( \tilde{G}_F^q(x,x^\prime)\ \d \hat{\sigma}_{qg\bar{q}}^--iG_F^q(x,x^\prime)\ \d \hat{\sigma}_{qg\bar{q}}^+ +\nn
 & & \hspace{-1.3cm}\tilde{G}_F^q(x^\prime,x^\prime-x)\ \d\hat{\sigma}_{q\bar{q}g}^- +iG_F^q(x^\prime,x^\prime-x)\ \d\hat{\sigma}_{q\bar{q}g}^+ +\mathrm{c.c.}\Big)\Big] .\nnn
\end{eqnarray}
Applying QCD lightcone gauge and working in a frame where the nucleon momentum $P$ and momentum transfer $q$ are collinear along the $z$-axis conveniently allows to derive Eq.~(\ref{eq:master}). In this formula the short- and long distance physics has been factorized into non-perturbative parton correlation functions and hard partonic cross sections $\d \hat{\sigma}$ as indicated in Fig.~\ref{DiagFact}. The latter are expressed through interferences of partonic amplitudes, i.e. $\d\hat{\sigma}_{q\bar{q}}\sim |M_q|^2$, $\d\hat{\sigma}_{qg\bar{q}}\sim M_{qg}M_q^\ast$ and $\d\hat{\sigma}_{q\bar{q}g}\sim M_{q\bar{q}}M_g^\ast$. As such $d\hat{\sigma}_{qg\bar{q}}$ and $d\hat{\sigma}_{q\bar{q}g}$ may carry real and imaginary parts. Kinematical approximations have been applied to the parton momenta to arrive at a collinear factorization formula (\ref{eq:master}), $k=xP+k_T$ for $\d \hat{\sigma}_{q\bar{q}}$, and $k=xP$, $p=x^\prime P$ for $\d \hat{\sigma}_{qg\bar{q}}$ and $\d \hat{\sigma}_{q\bar{q}g}$. In particular, parton virtualities are neglected in the hard factors. The superscript $\pm$ of the partonic cross sections in (\ref{eq:master}) refers to the quark helicity $\lambda_q$ that appears in the perturbative calculation of the hard factors $\d \hat{\sigma}$ by means of the quark projection $u_i(k,\lambda_q)\bar{u}_j(k,\lambda_q)=\tfrac{1}{2}[\slash{k}(1-\lambda_q \gamma_5)]_{ij}$. Thus, $\d\hat{\sigma}^\pm\equiv \d\hat{\sigma}^{\lambda_q=+1}\pm \d\hat{\sigma}^{\lambda_q=-1}$. 
The second line in Eq.~(\ref{eq:master}) describes 'kinematical twist-3' effects, i.e. effects induced by transverse quark motion. Hence, $\d\hat{\sigma}_{q\bar{q}}^{\partial,-}\sim (\partial \d \hat{\sigma}_{q\bar{q}}^-/\partial k_T^\rho)|_{k_T=0}$ and $\d\hat{\sigma}_{q\bar{q}}^{\partial,+}\sim (\partial \d \hat{\sigma}_{q\bar{q}}^+/\partial k_T^\rho)|_{k_T=0}$. The 'dynamical twist-3' effects are encoded in the last two lines of Eq.~(\ref{eq:master}). The partonic interference terms $\d\hat{\sigma}_{qg\bar{q}}^{\pm}$ and $\d\hat{\sigma}_{q\bar{q}g}^{\pm}$ provide information on the interaction of the lepton with more than one parton, and one may consider the $x^\prime$-integration in Eq.~(\ref{eq:master}) as 'phase space' in the initial state.\\
The soft parts in Eq.~(\ref{eq:master}) are parameterized in terms of the parton correlation functions $g_T$, $g_{1T}^{(1)}$, $f_{1T}^{\perp (1)}$, $G_F$ and $\tilde{G}_F$. For the precise definition in terms of hadronic matrix element the reader is referred to Ref.~\cite{ABMS}. The twist-3 quark correlation function $g_T$ generates the double spin observable $d\sigma_{LT}$ in the parton model. On the other hand the functions $f_{1T}^{\perp (1)}$ and $g_{1T}^{(1)}$ are $k_T$-moments of transverse momentum dependent parton distributions (TMDs) that describe the intrinsic transverse motion of quarks in a transversely polarized nucleon (cf.~\cite{ABMS,Baccetal}). The multipartonic quark-gluon correlation functions $G_F$ and $\tilde{G}_F$ (in the notation of Ref.~\cite{GFDef}) depend on two lightcone momentum fractions $x$ and $x^\prime$ of the quark and an additional gluon. Parity and hermiticity imply certain symmetry relations on $G_F$ and $\tilde{G}_F$, i.e. $G_F(x,x^\prime)=G_F(x^\prime,x)$ and $\tilde{G}_F(x,x^\prime)=-\tilde{G}_F(x^\prime,x)$. Consequently, $\tilde{G}_F(x,x)=0$. On the other hand, $G_F(x,x)$ can be non-zero. In fact, it is related to a $k_T$-moment of the Sivers function~\cite{BMPQS},
\begin{eqnarray}
f_{1T}^{\perp (1),q}(x)&=&\tfrac{\pi}{2}G_F^q(x,x)\ .\label{eq:Sivers}
\end{eqnarray}
Another constraint can be obtained from the QCD equation of motion which leads to a relation between the quark and quark-gluon correlation functions~\cite{Baccetal,ABMS} ($\mathrm{P}$ denotes the principal value prescription),
\begin{eqnarray}
&xg_T^q(x)-g_{1T}^{(1),q}(x)-\frac{m_q}{M}h_1^q(x) \ = \ x\tilde{g}_T^q(x)& \label{eq:EOM}\\
& = \mathrm{P}\int_0^1dx^\prime \frac{G_F^q(x,x^\prime)+\tilde{G}_F(x,x^\prime)}{2(x^\prime-x)}.&\nnn
\end{eqnarray}

{\it III. Perturbative calculations \---} 
\begin{figure}[t]
\psfig{file=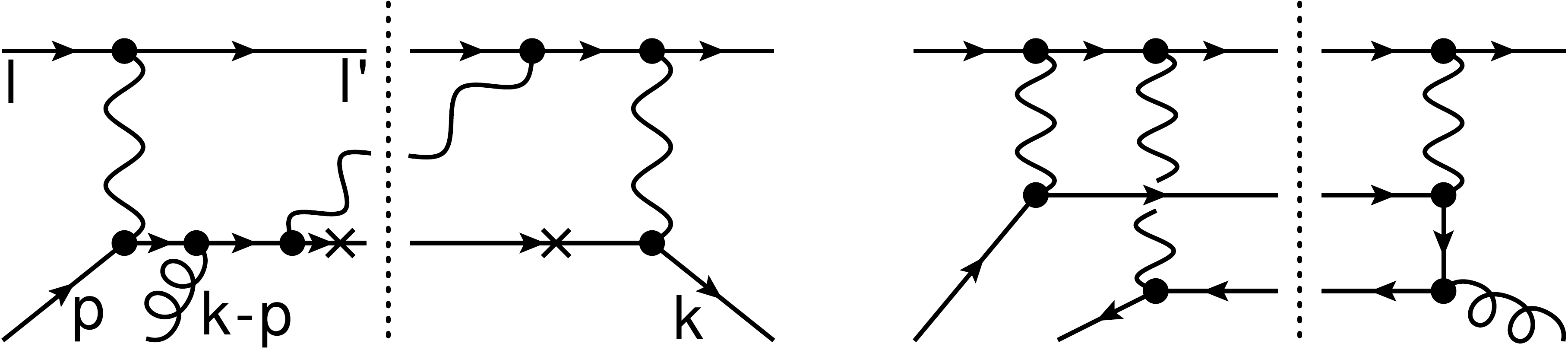,width=0.45\textwidth} 
\caption{{\it Left:} Sample diagrams with two unobserved particles in the final state contributing at $\mathcal{O}(e^3e_q^3)$ to $\d \hat{\sigma}_{q\bar{q}}^{\partial,+}$ (without non-perturbative gluon insertion) and $\d \hat{\sigma}_{qg\bar{q}}^{\pm}$ (with gluon insertion). The crosses indicate other possible gluon couplings. {\it Right:} Sample diagram contributing to $\d \hat{\sigma}_{q\bar{q}g}^{\pm}$.}
\label{real}
\end{figure}
It is illustrative to first consider a one-photon exchange in order to display the interplay of kinematical and dynamical twist-3 contributions. 
 Up to order $\mathcal{O}(\alpha_{\mathrm{em}}^2)$ only a tree-level diagram (left diagram of Fig.~\ref{real}, without photon emission) is relevant. Hence, calculating $\d\hat{\sigma}_{q\bar{q}}^{\perp}$, $\d\hat{\sigma}_{q\bar{q}}^{\partial,\pm}$ and $ \d\hat{\sigma}_{qg\bar{q}}^\pm$ to order $\mathcal{O}(\alpha_{\mathrm{em}}^2)$
leads to a simple result for the transverse target spin dependent cross section for an unpolarized {\it or} longitudinally polarized lepton,
\begin{eqnarray}
 \d\sigma_T^{1\gamma} & \propto & \sin\phi_s\sum_q e_q^2x_B (\tfrac{2}{\pi}f_{1T}^{(1),q}(x_B)-G_F^q(x_B,x_B)) \nn
 & &\hspace{-1.3cm}+\lambda \tfrac{x_B y}{2-y} \cos\phi_s \sum_q e_q^2(x_B g_{T}^q+g_{1T}^{(1),q}+x_B \tilde{g}_T^q)(x_B).\label{eq:OPE}
\end{eqnarray}
 Here, $\lambda$ denotes the helicity of a longitudinally polarized lepton. The relation (\ref{eq:Sivers}) implies that the single spin dependent cross section $\d\sigma_{UT}$ in the first line of (\ref{eq:OPE}) vanishes \--- as it should according to Ref.~\cite{Christ:1966zz}. Secondly, applying the relation (\ref{eq:EOM}) (with $m_q=0$) to the second line of Eq.~(\ref{eq:OPE}) reproduces a well-known result for the double spin dependent part, $\d\sigma_{LT}\propto g_T^q(x_B)$ (cf. Ref.~\cite{Baccetal}).

\begin{figure}[t]
\psfig{file=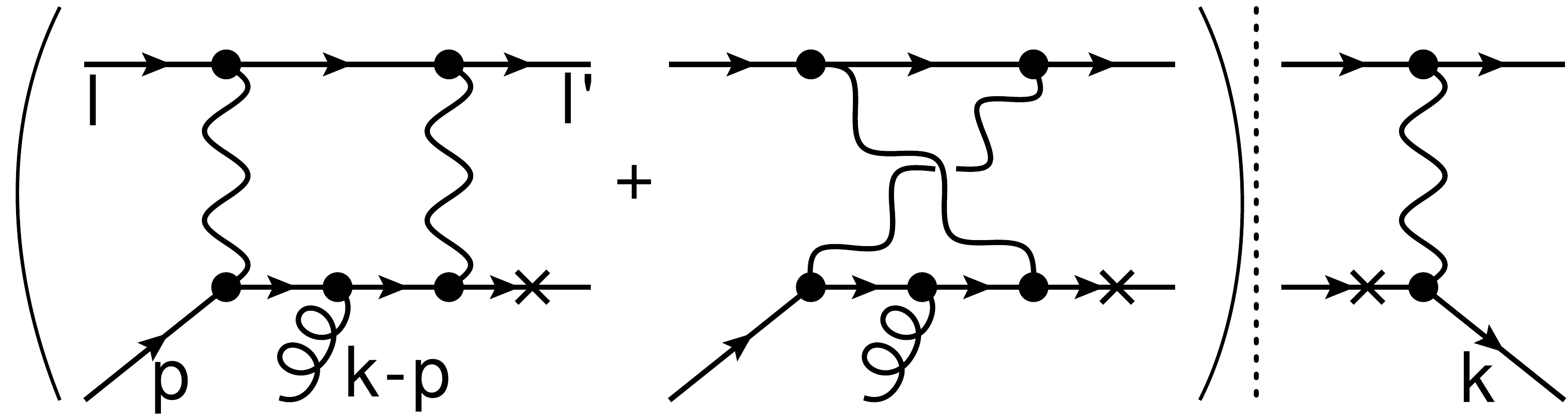,width=0.45\textwidth} 
\caption{ Virtual diagrams contributing at $\mathcal{O}(e^3e_q^3)$ to $\d \hat{\sigma}_{q\bar{q}}^{\perp}$, $\d \hat{\sigma}_{q\bar{q}}^{\partial,\pm}$ (without gluon insertion) and $\d \hat{\sigma}_{qg\bar{q}}^{\pm}$ (with gluon insertions, possible other gluon couplings indicated by the crosses).}
\label{virtual}
\end{figure}

A non-vanishing $\d\sigma_{UT}$ can only be derived from Eq.~(\ref{eq:master}) if the perturbative partonic cross sections $\d\hat{\sigma}$ are calculated to order $\mathcal{O}(\alpha_{\mathrm{em}}^3)$. The relevant diagrams are shown in Figs.~\ref{real} and \ref{virtual}. Those contributions determined by real photon emissions or $q\bar{q}$ creation as shown in Fig.~\ref{real} involve phase space integrals over two parton momenta. These integrals can be calculated in $4+2\varepsilon$ dimensions using the methods of Ref.~\cite{Neerven}. Since an imaginary part is necessary for a single spin asymmetry to occur certain propagators in Fig.~\ref{real} have to go onshell. This singles out three types of contributions: 'Soft Gluon Pole' (SGP) contributions proportional to $G_F^q(x,x)$, 'Soft Fermion Pole' (SFP) contributions proportional to $G_F^q(x,0)$, $\tilde{G}_F^q(x,0)$, and 'Hard Fermion Pole' (HFP) contributions proportional to $G_F^q(x,x_B)$, $\tilde{G}_F^q(x,x_B)$.\\
The SGP contributions from Fig.~\ref{real} ultimately cancel out if kinematical ($\d \hat{\sigma}_{q\bar{q}}^{\partial,+}$) and dynamical ($\d \hat{\sigma}_{qg\bar{q}}^{+}$) twist-3 partonic cross sections are combined via Eq.~(\ref{eq:Sivers}).  This observation results from partial integrations of the $x$- and $x^\prime$-integrals in Eq.~(\ref{eq:master}). Those partial integrations are necessary to get rid of derivative terms $\tfrac{\d}{\d x}G_F(x,x)$ that appear in intermediate steps. In this way the SGP contributions can be summed to the form $\int dx/x F^{\mathrm{SGP}}(x)\ G_F(x,x)$, with
\begin{equation}
  F^{\mathrm{SGP}}=\tfrac{1}{2}\d \hat{\sigma}_{q\bar{q}}^{\partial,+}-\tfrac{x}{2}\tfrac{d}{dx}\d \hat{\sigma}_{qg\bar{q}}^+(x,x)+x \tfrac{\partial}{\partial x^\prime}\d \hat{\sigma}_{qg\bar{q}}^+(x,x^\prime)\big|_{x\prime=x}.\nonumber
\end{equation}
The combination $F^{\mathrm{SGP}}$ vanishes to order $\mathcal{O}(\varepsilon^0)$.\\
Soft Fermionic Poles appear in the interference terms $\d \hat{\sigma}_{qg\bar{q}}^{\pm}(x,0)$ and $\d \hat{\sigma}_{q\bar{q}g}^{\pm}(x,x)$, i.e. in the left and right diagrams of Fig.~\ref{real}. One observes cancellations between those diagrams for $\d \hat{\sigma}_{qg\bar{q}}^{\pm}(x,0)$ and $\d \hat{\sigma}_{q\bar{q}g}^{\pm}(x,x)$ such that Soft Fermionic Poles do not contribute at $\mathcal{O}(\varepsilon^0)$ to $\d \sigma_{UT}$ in Eq.~(\ref{eq:master}).\\
Hard Fermionic Poles are encoded in the photon emission diagrams for $\d \hat{\sigma}_{qg\bar{q}}^\pm$ on the left of Fig.~\ref{real}, for $x^\prime=x_B$. They do not cancel and can be expressed as follows,
\begin{eqnarray}
\d\sigma^{\mathrm{HFP}} _{UT}& \propto & \int_{x_B}^1\d x \frac{\hat{C}^{\mathrm{HFP}}_+\ G_F(x,x_B)+\hat{C}^{\mathrm{HFP}}_-\ \tilde{G}_F(x,x_B)}{(x-x_B)_+}\nn
 & & +R^{\mathrm{HFP}}(x_B)G_F(x_B,x_B).\label{eq:HFP}
\end{eqnarray}
The coefficient functions $\hat{C}^{\mathrm{HFP}}_\pm (x)$ are finite as $\varepsilon \to 0$, and the integral is well-defined at the endpoint $x\to x_B$ by means of the plus-prescription $(...)_+$ (defined in, e.g., Ref.~\cite{ABMS}). The remainder $R^\mathrm{HFP}$ carries a $1/\varepsilon$ pole indicating a soft singularity at $x=x^\prime=x_B$.

The two-photon exchange (TPE) contributions to the partonic cross sections $\d \hat{\sigma}_{q\bar{q}}^{\perp}$, $\d \hat{\sigma}_{q\bar{q}}^{\partial \pm}$, $\d \hat{\sigma}_{qg\bar{q}}^\pm$ are shown in Fig.~\ref{virtual}. They are determined by virtual one-loop diagrams that can be calculated using the methods of Ref.~\cite{Bern}. Since only one unobserved quark enters the final state for those diagrams the phase space integration becomes trivial and enforces $x=x_B$. Both real- and imaginary parts of the loops are relevant for the transverse target SSA in DIS.

The imaginary part of the loops in Fig.~\ref{virtual} without gluon insertions generate the 'kinematical twist-3' terms $\d \hat{\sigma}_{q\bar{q}}^{\perp}$ and $\d \hat{\sigma}_{q\bar{q}}^{\partial -}$ in Eq.~(\ref{eq:master}). 
They were shown to be equal in Ref.~\cite{OwnWork1}, i.e. $\d \hat{\sigma}_{q\bar{q},\mathrm{TPE}}^{\perp}=\d \hat{\sigma}_{q\bar{q},\mathrm{TPE}}^{\partial -}$. Hence, the correlation functions $g_T$ and $g_{1T}^{(1)}$ in (\ref{eq:master}) can be combined to $xg_T-g_{1T}^{(1)}$. 
By means of the relation (\ref{eq:EOM}) this term is expressed through quark-gluon correlation functions $G_F$ and $\tilde{G}_F$. The result of this manipulation can then be added to the 'dynamical twist-3' TPE terms $\d \hat{\sigma}_{qg\bar{q},\mathrm{TPE}}^{\pm}$ in the third line of Eq.~(\ref{eq:master}) which are generated by the imaginary part of the loops with gluon insertion in Fig.~\ref{virtual}. Eventually the TPE contributions read,
\begin{eqnarray}
\d\sigma^{\mathrm{TPE}}_{UT}  & \propto & \int_{x_B}^1\d x^\prime \frac{\hat{C}^{\mathrm{TPE}}_+\ G_F(x_B,x^\prime)+\hat{C}^{\mathrm{TPE}}_-\ \tilde{G}_F(x_B,x^\prime)}{(x^\prime-x_B)_+} \nn
 & &\hspace{-0.6cm}+\mathrm{P}\int_{0}^1\d x^\prime \frac{\hat{D}^{\mathrm{TPE}}_+\ G_F(x_B,x^\prime)+\hat{D}^{\mathrm{TPE}}_-\ \tilde{G}_F(x_B,x^\prime)}{x^\prime-x_B} \nn
 & &+ R^{\mathrm{TPE}}(x_B)G_F(x_B,x_B).\label{eq:TPE}
\end{eqnarray}
Again, the coefficient functions $\hat{C}^{\mathrm{TPE}}_{\pm}(x^\prime)$ and  $\hat{D}^{\mathrm{TPE}}_{\pm}(x^\prime)$ are finite as $\varepsilon\to 0$ while the remainder $R^{\mathrm{TPE}}$ is singular due to a $1/\varepsilon$ pole. 

The real parts of the TPE loops in Fig.~\ref{virtual} contribute to $\d \hat{\sigma}_{q\bar{q}}^{\partial +}$ and $\d \hat{\sigma}_{qg\bar{q}}^{+}$ at $x^\prime=x=x_B$. Both cross section can be added using the relation~(\ref{eq:Sivers}). In this sum $1/\varepsilon^2$ poles cancel while the remainder, 
$R^{\mathrm{real}}(x_B)=\tfrac{1}{2}\d \hat{\sigma}_{q\bar{q},\mathrm{real}}^{\partial,+}(x_B)-\d \hat{\sigma}_{qg\bar{q},\mathrm{real}}^{+}(x_B,x_B)$, carries a $1/\varepsilon$ pole. All $1/\varepsilon$ poles then eventually cancel in the sum of the three remainders,
$R^{\mathrm{HFP}}+R^{\mathrm{TPE}}+R^{\mathrm{real}}$.

As a result, summing Eqs.~(\ref{eq:HFP}), (\ref{eq:TPE}) and $R^{\mathrm{real}}$ leads to a well-behaved formula. 
Defining some auxilliary functions with $\alpha=x/x_B$ and $f(y)=1-y+\tfrac{y^2}{2}$,
\begin{eqnarray}
d(\alpha,y) & = & \tfrac{\tfrac{1}{2}(1-y)(1+2y-y^2+\alpha(3-4y+y^2))}{y+\alpha (1-y)},\label{eq:aux}\\
e(\alpha,y) & = & \tfrac{\tfrac{1}{2}(1-y)(-1+2y+y^2+\alpha(1-y^2))}{y+\alpha(1-y)},\nn
g(\alpha,y) & = & \tfrac{\ln y}{\alpha}+\tfrac{\ln (y+\alpha(1-y))}{\alpha(\alpha-1)},\nn
h(\alpha,y) & = & \ln y+(1-y)^2\tfrac{\ln \alpha}{\alpha-1},\nn
\hat{C}_+(x,x_B,y)& = & \delta(x-x_B)\big(f(y)\ln(\tfrac{1-x}{x})-(1-\tfrac{y}{2})^2\big)\nn
 & & +\tfrac{\theta(x-x_B)}{(x-x_B)_+}f(y)+\mathrm{P}\tfrac{h(\alpha,y)+g(\alpha,y)+d(\alpha,y)}{x-x_B},\nn
\hat{C}_-(x,x_B,y)& = &\tfrac{\theta(x-x_B)}{(x-x_B)_+}f(y)+\mathrm{P}\tfrac{h(\alpha,y)-g(\alpha,y)+e(\alpha,y)}{x-x_B},\nnn
\end{eqnarray}
the complete single transverse dependent DIS cross section at $\mathcal{O}(\alpha_{\mathrm{em}}^3)$ reads,
\begin{eqnarray}
\d\sigma_{UT} & = & -|S_T|\sin\phi_s \frac{4\alpha_{\mathrm{em}}^3}{yQ^4} \frac{M}{Q}\frac{x_B y}{\sqrt{1-y}}\sum_q\times\label{eq:result}\\
& & \hspace{-1cm} \Big[ e_q^3\int_0^1dx\Big(\hat{C}_+\ G_F^q(x_B,x)+\hat{C}_-\ \tilde{G}_F^q(x_B,x)\Big)+\nn
& & \hspace{-1.4cm}(1-y)\tfrac{e_q^3 m_q}{M}h_1^q(x_B)+\tfrac{2-y}{2y}e_q^2(1-x_B\tfrac{\d}{\d x_B}) G_F^{\gamma,q}(x_B,x_B)\Big].\nnn
\end{eqnarray}
The finite quark mass term of Ref.~\cite{Afanasev,new} has been added as well as the contribution of Ref.~\cite{new} describing scenario (ii) where the two photons couple to different quarks. The latter term involves a quark-photon correlation function $G_F^\gamma$. Notice a slight redefinition $G_F^\gamma(x,x)\equiv\tfrac{1}{2e^2}F_{FT}(x,x)$ of the object $F_{FT}$ introduced in \cite{new}.
A theoretical formula for the SSA $A_{UT}=(d\sigma_{UT}(\phi_s)-d\sigma_{UT}(\phi_s-\pi))/(2d\sigma_{UU})$ can be immediately obtained from Eq.~(\ref{eq:result}) by dividing by the well-known parton model result for the unpolarized cross section~\cite{Baccetal},
$\d\sigma_{UU}  = \tfrac{4\alpha_{\mathrm{em}}^2}{Q^4y}f(y)\sum_qe_q^2x_B f_1^q(x_B)$, with $f_1$ the unpolarized collinear parton distribution.\\
{\it IV. Conclusions-} The formula (\ref{eq:result}) for the single transverse spin dependent cross section, $\d\sigma_{UT}$, derived in a partonic picture, is the main result of this paper. It suggests that precision measurements of the $x_B$-dependence as well as the $y$-dependence of $A_{UT}$ may help to constrain multiparton correlations in the nucleon. In particular the full support of quark-gluon correlation functions can be accessed experimentally through measurements of this observable. Interestingly, very recently an alternative way to probe the support of $G_F(x,x^\prime)$ and $\tilde{G}_F(x,x^\prime)$ for $x\neq x^\prime$ was suggested in double polarized proton collisions at RHIC~\cite{ppLT}. A combined analysis of potential future DIS and $pp$ data would allow to test the evolution and universality of quark-gluon correlation functions.

\begin{acknowledgments}
I thank Andreas Metz and Werner Vogelsang for valuable discussions on this subject.
\end{acknowledgments}

\end{document}